\documentclass[12pt]{article}
\usepackage{graphicx}
\usepackage{lscape}
\usepackage{cite}
\usepackage{amssymb}
\usepackage{appendix}
\usepackage{multirow}
\usepackage{subfigure}

\begin{document}
\def\qq{\langle \bar q q \rangle}
\def\uu{\langle \bar u u \rangle}
\def\dd{\langle \bar d d \rangle}
\def\sp{\langle \bar s s \rangle}
\def\GG{\langle g_s^2 G^2 \rangle}
\def\Tr{\mbox{Tr}}
\def\figt#1#2#3{
        \begin{figure}
        $\left. \right.$
        \vspace*{-2cm}
        \begin{center}
        \includegraphics[width=10cm]{#1}
        \end{center}
        \vspace*{-0.2cm}
        \caption{#3}
        \label{#2}
        \end{figure}
    }

\def\figb#1#2#3{
        \begin{figure}
        $\left. \right.$
        \vspace*{-1cm}
        \begin{center}
        \includegraphics[width=10cm]{#1}
        \end{center}
        \vspace*{-0.2cm}
        \caption{#3}
        \label{#2}
        \end{figure}
                }

\def\ds{\displaystyle}
\def\beq{\begin{equation}}
\def\eeq{\end{equation}}
\def\bea{\begin{eqnarray}}
\def\eea{\end{eqnarray}}
\def\beeq{\begin{eqnarray}}
\def\eeeq{\end{eqnarray}}
\def\ve{\vert}
\def\vel{\left|}
\def\ver{\right|}
\def\nnb{\nonumber}
\def\ga{\left(}
\def\dr{\right)}
\def\aga{\left\{}
\def\adr{\right\}}
\def\lla{\left<}
\def\rra{\right>}
\def\rar{\rightarrow}
\def\lrar{\leftrightarrow}
\def\nnb{\nonumber}
\def\la{\langle}
\def\ra{\rangle}
\def\ba{\begin{array}}
\def\ea{\end{array}}
\def\tr{\mbox{Tr}}
\def\ssp{{\Sigma^{*+}}}
\def\sso{{\Sigma^{*0}}}
\def\ssm{{\Sigma^{*-}}}
\def\xis0{{\Xi^{*0}}}
\def\xism{{\Xi^{*-}}}
\def\qs{\la \bar s s \ra}
\def\qu{\la \bar u u \ra}
\def\qd{\la \bar d d \ra}
\def\qq{\la \bar q q \ra}
\def\gGgG{\la g^2 G^2 \ra}
\def\q{\gamma_5 \not\!q}
\def\x{\gamma_5 \not\!x}
\def\g5{\gamma_5}
\def\sb{S_Q^{cf}}
\def\sd{S_d^{be}}
\def\su{S_u^{ad}}
\def\sbp{{S}_Q^{'cf}}
\def\sdp{{S}_d^{'be}}
\def\sup{{S}_u^{'ad}}
\def\ssp{{S}_s^{'??}}

\def\sig{\sigma_{\mu \nu} \gamma_5 p^\mu q^\nu}
\def\fo{f_0(\frac{s_0}{M^2})}
\def\ffi{f_1(\frac{s_0}{M^2})}
\def\fii{f_2(\frac{s_0}{M^2})}
\def\O{{\cal O}}
\def\sl{{\Sigma^0 \Lambda}}
\def\es{\!\!\! &=& \!\!\!}
\def\ap{\!\!\! &\approx& \!\!\!}
\def\md{\!\!\!\! &\mid& \!\!\!\!}
\def\ar{&+& \!\!\!}
\def\ek{&-& \!\!\!}
\def\kek{\!\!\!&-& \!\!\!}
\def\cp{&\times& \!\!\!}
\def\se{\!\!\! &\simeq& \!\!\!}
\def\eqv{&\equiv& \!\!\!}
\def\kpm{&\pm& \!\!\!}
\def\kmp{&\mp& \!\!\!}
\def\mcdot{\!\cdot\!}
\def\erar{&\rightarrow&}

% .........................................................

\def\simlt{\stackrel{<}{{}_\sim}}
\def\simgt{\stackrel{>}{{}_\sim}}

% .........................................................

\title{
         {\Large
                 {\bf
                     Semileptonic transition of $\Sigma_{b}$ to $\Sigma$ in Light Cone QCD Sum Rules                  }
         }
      }

\author{\vspace{1cm}\\
{\small K. Azizi$^a$ \thanks {e-mail: kazizi@dogus.edu.tr}\,, M.
Bayar$^{b,c}$ \thanks {e-mail: melahat.bayar@kocaeli.edu.tr}\,\,, A.
Ozpineci$^d$ \thanks {e-mail: ozpineci@metu.edu.tr}\,, Y. Sarac$^e$
\thanks {e-mail: ysoymak@atilim.edu.tr}\,\,, H.
Sundu$^b$ \thanks {e-mail: hayriye.sundu@kocaeli.edu.tr}} \\
{\small $^a$  Physics Department,  Do\u gu\c s University,
 Ac{\i}badem-Kad{\i}k\"oy, 34722 Istanbul, Turkey} \\
{\small $^b$ Department of Physics, Kocaeli University, 41380 Izmit, Turkey} \\
{\small $^c$Instituto de F{\'\i}sica Corpuscular (centro mixto CSIC-UV), }\\{\small Institutos de Investigaci\'on de Paterna, Aptdo. 22085, 46071, Valencia, Spain} \\
{\small $^d$ Physics
Department, Middle East Technical University, 06531, Ankara, Turkey} \\
{\small $^e$Electrical and Electronics Engineering Department,
Atilim University, 06836 Ankara, Turkey} }
\date{}

\begin{titlepage}
\maketitle
\thispagestyle{empty}

\begin{abstract}
We use  distribution amplitudes of the light $\Sigma$ baryon and the most general form of the interpolating current for heavy $\Sigma_b$ baryon to investigate the semileptonic
$\Sigma_b\rightarrow \Sigma l^+l^-$ transition in light cone QCD sum rules. We calculate all twelve form factors responsible for this transition and use them to evaluate the branching ratio of the considered channel.
The order of branching fraction shows that this channel can be detected at LHC.

\end{abstract}

~~~PACS number(s): 11.55.Hx, 13.30.-a, 13.30.Ce, 14.20.Mr
\end{titlepage}

%%%
\section{Introduction}

The systems involving heavy quarks decays are important frameworks
to restrict the standard model (SM) parameters as well as search for
new physics effects. Especially, the flavor changing neutral current
(FCNC) transition of $b \rightarrow s \bar \ell \ell$, which is
underlying transition of $\Sigma_b\rightarrow \Sigma l^+l^-$ decay
at the quark level, is known to be sensitive to new physics effects.
 This process can also be used in exact determination of the   $V_{tb}$ and $V_{ts}$ as elements of the Cabibbo-Kobayashi-Maskawa (CKM) matrix  and answering some fundamental questions such as CP
violation.

In the last decade, important experimental progress has been made in
identification and spectroscopy of the heavy baryons with single
heavy  quark
\cite{Mattson,Ocherashvili,Acosta,Chistov,Aubert1,Abazov1,Aaltonen1,Solovieva}.
It is expected that the LHC will open new horizons  not only in the
identification and spectroscopy of these baryons, but also it will
provide possibility to study the weak, strong and electromagnetic
decays of heavy baryons.

In accordance with this experimental progress, there is an increasing interest
 on calculation of  parameters of the heavy baryons and investigation of their decay modes theoretically.
 The masses of these baryons have been calculated using various
methods such as quark models
\cite{Ebert1,Ebert3,Capstick,Matrasulov,Gershtein,Kiselev3,Vijande,Martynenko,Hasenfratz},
heavy quark effective theory
\cite{Grozin,Groote,Dai,Lee,Huang,Liu1,Wang2} and QCD sum rules
\cite{kazem1,Shuryak,Kiselev1,Kiselev2,Bagan1,Bagan3,Duraes,Wang1,Zhang1}.
Besides the mass spectrum, their weak, strong and electromagnetic
decays have also received special attention, recently (for instance
see
\cite{ebert3,Ming-Qiu,Albertus,Flores-Mendieta,Pervin,Azizi-Bayar2,Azizi-Bayar3,Azizi-Sundu,boyuk1,boyuk2}
and references therein).

In the present work, we analyze the semileptonic
$\Sigma_b\rightarrow\Sigma l^+l^-$ transition in the framework of the light cone QCD sum rules.
The main ingredients in analysis of this channel are form factors entering the transition matrix elements.
Using the most general form of the interpolating field for the $\Sigma_b$ heavy baryon as well as the distribution amplitudes (DA's) of the light $\Sigma$ baryon, we first calculate all twelve form factors in full theory.
Then, we use these form factors to calculate the total decay rate as well as the branching ratio of the considered decay channel.

The paper is organized in three sections. In the next
section, we obtain QCD sum rules for the form factors. In section 3, we  numerically  analyze the form factors and use them   to calculate the  related decay rate and branching fraction.

%%%
\section{light cone QCD sum rules for form factors}

In this section, we focus on the calculation of the form factors
corresponding to  $\Sigma_b\rightarrow\Sigma l^+l^-$ semileptonic
decay which proceeds via $b\rightarrow s$ transition at quark level.
The effective Hamiltonian  describing this transition is written as:
\begin{eqnarray} \label{ham} {\cal H}_{eff} \es \frac{G_F~\alpha_{em}
V_{tb}~V_{ts}^{^{*}}}{2\sqrt2~\pi} \Bigg\{\vphantom{\int_0^{x_2}}
C_{9}^{eff}~ \bar{s} \gamma_\mu (1-\gamma_5) b \bar l \gamma^\mu
 l +C_{10} ~\bar{s}
\gamma_\mu (1-\gamma_5) b \bar l \gamma^\mu \gamma_{5}l \nnb \\
\ek2 m_{b}~C_{7}\frac{1}{q^{2}} ~\bar{s} i \sigma_{\mu\nu}q^{\nu}
(1+\gamma_5) b \bar l \gamma^\mu l \Bigg\}~.
\end{eqnarray}
The amplitude of  the transition can be obtained by sandwiching the effective Hamiltonian between the initial and final states,
\begin{eqnarray} \label{matrix}
 {\cal M}  \es \frac{G_F~\alpha_{em}
V_{tb}~V_{ts}^{^{*}}}{2\sqrt2~\pi} \Bigg\{\vphantom{\int_0^{x_2}}
C_{9}^{eff}~ \langle \Sigma \vert \bar{s} \gamma_\mu (1-\gamma_5) b \vert \Sigma_b \rangle \bar l \gamma^\mu
 l +C_{10} ~\langle \Sigma \vert \bar{s}
\gamma_\mu (1-\gamma_5) b \vert \Sigma_b \rangle \bar l \gamma^\mu \gamma_{5}l \nnb \\
\ek2 m_{b}~C_{7}\frac{1}{q^{2}} ~\langle \Sigma \vert \bar{s} i \sigma_{\mu\nu}q^{\nu}
(1+\gamma_5) b \vert \Sigma_b \rangle \bar l \gamma^\mu l \Bigg\}~.
\end{eqnarray}
From this equation, it is obvious  that  the transition matrix elements
$\langle \Sigma(p) \vert \bar{s} \gamma_\mu (1-\gamma_5) b \vert \Sigma_b(p+q)$
and  $\langle \Sigma(p) \vert \bar{s} \sigma_{\mu \nu} q^\nu (1+\gamma_5) b \vert \Sigma_b(p+q) \rangle$ are required.
These matrix elements are expressed in terms of  twelve form factors $f_i$, $g_i$, $f^T_i$ and $g^T_i$ ($i$ running from 1 to 3) as follows:
\begin{eqnarray}\label{matrixel1a} \langle \Sigma(p) \md  \bar s \gamma_\mu
(1-\gamma_5) b \mid \Sigma_b(p+q)\rangle= \bar {u}_\Sigma(p)
\Big[\gamma_{\mu}f_{1}(q^{2})+{i}
\sigma_{\mu\nu}q^{\nu}f_{2}(q^{2}) + q^{\mu}f_{3}(q^{2}) \nnb \\
\ek \gamma_{\mu}\gamma_5
g_{1}(q^{2})-{i}\sigma_{\mu\nu}\gamma_5q^{\nu}g_{2}(q^{2}) -
q^{\mu}\gamma_5 g_{3}(q^{2}) \vphantom{\int_0^{x_2}}\Big]
u_{\Sigma_{b}}(p+q)~,
\end{eqnarray}
and
\begin{eqnarray}\label{matrixel1b} \langle \Sigma(p)\md \bar s i
\sigma_{\mu\nu}q^{\nu} (1+ \gamma_5) b \mid \Sigma_b(p+q)\rangle
=\bar{u}_\Sigma(p)
\Big[\gamma_{\mu}f_{1}^{T}(q^{2})+{i}\sigma_{\mu\nu}q^{\nu}f_{2}^{T}(q^{2})+
q^{\mu}f_{3}^{T}(q^{2}) \nnb \\
\ar \gamma_{\mu}\gamma_5
g_{1}^{T}(q^{2})+{i}\sigma_{\mu\nu}\gamma_5q^{\nu}g_{2}^{T}(q^{2}) +
q^{\mu}\gamma_5 g_{3}^{T}(q^{2}) \vphantom{\int_0^{x_2}}\Big]
u_{\Sigma_{b}}(p+q)~,
\end{eqnarray}
where  $u_{\Sigma_b}$ and $u_{\Sigma}$ are the spinors
of $\Sigma_b$ and $\Sigma$ baryons, respectively, and   $q$ denotes  transferred
momentum.

Our main task is to calculate the aforesaid transition form factors. In accordance with
the philosophy of QCD sum rules, we start considering the following
correlation functions:
\begin{eqnarray}\label{T} \Pi^I_{\mu}(p,q) = i\int d^{4}xe^{-iqx}\langle 0 \mid
T\{ J^{\Sigma_{b}}(0), \bar b(x) \gamma_\mu (1-\gamma_5) s(x))\}\mid
\Sigma(p)\rangle~,\nnb\\
\Pi^{II}_{\mu}(p,q) = i\int d^{4}xe^{-iqx}\langle 0 \mid T\{
J^{\Sigma_{b}}(0),\bar{b}(x) i \sigma_{\mu\nu}q^{\nu} (1+ \gamma_5)
s(x)\}\mid \Sigma(p)\rangle~,
\end{eqnarray}
 where $J^{\Sigma_b}$ stands for interpolating
current of $\Sigma_b$. The interpolating current should be chosen as a composite operator that has the same quantum numbers as the baryon under study. For $\Sigma_b$ baryon,
there are two possible choices for such a current that does not contain any derivatives or auxiliary four vectors. The most general form for the interpolating current is a superposition of these two choices. Hence, for the interpolating current of the $\Sigma_b$ baryon,  the operator
\begin{eqnarray}\label{cur.Sigma}
J^{\Sigma_{b}}(x)\es\frac{-1}{\sqrt{2}}\epsilon_{abc}
\Bigg\{\vphantom{\int_0^{x_2}}[u_{1}^{aT}(x)Cb^{b}(x)]
\gamma_{5}d^{c}(x)
+\beta[u_{1}^{aT}(x)C\gamma_{5}b^{b}(x)]d^{c}(x) \nnb \\
\ek [b^{aT}(x)Cd^{b}(x)]\gamma_{5}u^{c}(x) -\vphantom{\int_0^{x_2}}
\beta[b^{aT}(x)C\gamma_{5}d^{b}(x)]u^{c}(x) \Bigg\}~,
\end{eqnarray}
is chosen. Here $C$ is the charge conjugation operator, $\beta$ is an
arbitrary parameter, and   $a$, $b$, and $c$, are
the color indices. Taking $\beta=-1$ corresponds to the Ioffe current.

The correlation function can be calculated both in terms of the hadronic parameters, such as the form factors, and also in terms of the QCD parameters.
The expression in terms of the QCD parameters is evaluated by expanding the time
ordered product of the  currents in
terms of the $\Sigma$ distribution amplitudes via operator product
expansion (OPE) in deep Euclidean region. On the other hand, the
physical counterpart is calculated by inserting a complete set of intermediate states.
The two expression are then matched using dispersion relations.
%After the above procedure the physical
%quantity that we deal with is obtained through matching the
%theoretical side and physical side of the calculation via the
%dispersion relation.

To begin with, let us evaluate the correlation  function in terms of hadronic parameters. After inserting
the complete set of intermediate states into the correlation functions
and isolating the ground state contribution we obtain
\begin{eqnarray} \label{phys1} \Pi_{\mu}^{I}(p,q)=\sum_{s}\frac{\langle 0\mid
J^{\Sigma_{b}}(0) \mid \Sigma_{b}(p+q,s)\rangle\langle
\Sigma_{b}(p+q,s)\mid  \bar b \gamma_\mu (1-\gamma_5) s \mid
\Sigma(p)\rangle}{m_{\Sigma_{b}}^{2}-(p+q)^{2}}+\cdots~,
\end{eqnarray}
\begin{eqnarray}
\label{phys1111} \Pi_{\mu}^{II}(p,q)=\sum_{s}\frac{\langle 0\mid
J^{\Sigma_{b}}(0) \mid \Sigma_{b}(p+q,s)\rangle\langle
\Sigma_{b}(p+q,s)\mid \bar{b} i \sigma_{\mu\nu}q^{\nu} (1+ \gamma_5)
s \mid \Sigma(p)\rangle}{m_{\Sigma_{b}}^{2}-(p+q)^{2}}+\cdots~,
\end{eqnarray}
where the $\cdots$ stands for  contributions of the higher states
and continuum, and the sum is over the polarizations of the $\Sigma_b$ baryon. The matrix element of the interpolating current between the vacuum and the $\Sigma_b$ baryon  appearing in  Eqs.~(\ref{phys1})
and (\ref{phys1111}), $\langle0\mid J^{\Sigma_{b}}(0)\mid
\Sigma_{b}(p+q,s) \rangle$, can be expressed in terms of the residue of the $\Sigma_b$ baryon defined
as:
\begin{eqnarray}\label{matrixel2} \langle0\mid J^{\Sigma_{b}}(0)\mid
\Sigma_{b}(p+q,s)\rangle=\lambda_{\Sigma_{b}}
u_{\Sigma_{b}}(p+q,s)~.
\end{eqnarray}
The other matrix elements in Eqs.~(\ref{phys1}) and
(\ref{phys1111}) are defined in terms of the form
factors as previously shown.
Combining Eqs.~(\ref{matrixel1a}),
(\ref{matrixel1b}), and (\ref{phys1})--(\ref{matrixel2}) and summing over the polarization of the $\Sigma_b$ baryon using the  expression
\begin{eqnarray}\label{spinor}
\sum_{s}u_{\Sigma_{b}}(p+q,s)\overline{u}_{\Sigma_{b}}(p+q,s)=
\not\!p+\not\!q+m_{\Sigma_{b}}~,
\end{eqnarray}
the correlation functions can be expressed as:
\begin{eqnarray}\label{phys2} \Pi_{\mu}^{I}(p,q)\es
\lambda_{\Sigma_{b}} {\not\!p +\not\!q +m_{\Sigma_{b}}\over
m_{\Sigma_{b}}^{2}-(p+q)^{2}} \Big\{\gamma_{\mu}f_{1} -
i\sigma_{\mu\nu} q^{\nu} f_{2} +
q_{\mu} f_{3} \nnb \\
&&- \gamma_{\mu}\gamma_5 g_{1} - i\sigma_{\mu\nu} q^{\nu} \gamma_5
g_{2} +q_{\mu}\gamma_5 g_{3}
\vphantom{\int_0^{x_2}}\Big\} u_\Sigma(p) ~, \\ \nnb \\
\label{phys22} \Pi_{\mu}^{II}(p,q)\es \lambda_{\Sigma_{b}}{\not\!p
+\not\!q +m_{\Sigma_b}\over m_{\Sigma_{b}}^{2}-(p+q)^{2}}
\Big\{\gamma_{\mu}f_{1}^{T}- i \sigma_{\mu\nu}q^{\nu}f_{2}^{T} +
q_{\mu}f_{3}^{T}  \nnb \\
\ar \gamma_{\mu}\gamma_5 g_{1}^{T} + i
\sigma_{\mu\nu}\gamma_5q^{\nu}g_{2}^{T} - q_{\mu}\gamma_5 g_{3}^{T}
\Big\} u_\Sigma(p)~.
\end{eqnarray}
Commuting $\not\!p$ all the way to the right and using the equation of motion to write $\not\!p u_\Sigma(p) = m_\Sigma u_\Sigma(p)$,  Eqs.~(\ref{phys2}) and
(\ref{phys22}) lead to the final expressions for the phenomenological side:
\begin{eqnarray}\label{sigmaaftera} \Pi_{\mu}^{I}(p,q)\es
\frac{\lambda_{\Sigma_{b}}}{m_{\Sigma_{b}}^{2}-(p+q)^{2}} \Big\{ 2
f_{1}(q^{2})p_\mu + 2 f_{2}(q^{2}) p_\mu\not\!q
+\Big[ f_2(q^2) + f_3(q^2)\Big] q_\mu\not\!q \nnb \\
\ek 2 g_1(q^2) p_{\mu}\gamma_5 + 2 g_2(q^2)p_\mu\not\!q\gamma_5 +
\Big[g_2(q^2)+g_3(q^2)\vphantom{\int_0^{x_2}}\Big] q_\mu\not\!q
\gamma_5 \nnb \\
\ar \mbox{\rm other structures}\Big\} u_\Sigma(p)~, \\ \nnb \\
\label{sigmaafterb} \Pi_{\mu}^{II}(p,q)\es
\frac{\lambda_{\Sigma_{b}}}{m_{\Sigma_{b}}^{2}-(p+q)^{2}} \Big\{ 2
f_{1}^{T}(q^{2}) p_\mu + 2 f_{2}^{T}(q^{2})p_\mu\not\!q+
\Big[ f_2^{T}(q^2) + f_3^{T}(q^2)\Big]q_\mu\not\!q \nnb \\
\ar 2g_1^{T}(q^2)p_{\mu}\gamma_5 - 2
g_2^{T}(q^2)p_\mu\not\!q\gamma_5
-\Big[g_2^{T}(q^2)+g_3^{T}(q^2) \Big]q_\mu\not\!q\gamma_5 \nnb \\
\ar \mbox{\rm other structures} \Big\} u_\Sigma(p)~.
\end{eqnarray}
In these two expressions only the independent structures,
$p_{\mu}$, $p_{\mu}\rlap/q$, $q_{\mu}\rlap/q$, $p_{\mu}\gamma_5$,
$p_{\mu}\rlap/q\gamma_5$, and $q_{\mu}\rlap/q \gamma_5$, are
presented explicitly, owing to their sufficiency to determine the
aimed form factors, $f_{1}(f_{1}^{T})$, $f_{2}(f_{2}^{T})$,
$f_{2}+f_{3}(f_{2}^{T}+f_{3}^{T})$, $g_{1}(g_{1}^{T})$,
$g_{2}(g_{2}^{T})$ and $g_{2}+g_{3}(g_{2}^{T}+g_{3}^{T})$.

After completing the evaluation of the correlation function in terms of the hadronic parameters, now let us focus our
attention on evaluating the correlation function in terms of the QCD parameters and the DA's of the $\Sigma$ baryon . After placing the explicit
expression of interpolating current given in Eq.~(\ref{cur.Sigma})
into Eq.~(\ref{T}) and contracting out the heavy quark operators, we attain the following representation of the
correlators in QCD side:
\begin{eqnarray}\label{mut.m}
\Pi^I_{\mu} &=& \frac{-i}{\sqrt{2}} \epsilon^{abc}\int d^4x e^{-iqx}
\Bigg\{\Bigg(\Big[( C )_{\eta\beta} (\gamma_5)_{\rho\phi}-( C
)_{\beta\phi} (\gamma_5)_{\rho\eta}\Big]  +\beta\Bigg[(C \gamma_5
)_{\eta\beta}(I)_{\rho\phi}
 \nonumber \\
&-& (C \gamma_5 )_{\beta\phi}(I)_{\rho\eta} \Bigg]\Bigg) \Big[
\gamma_{\mu}(1-\gamma_5)
\Big]_{\sigma\theta}\Bigg\}S_Q(-x)_{\beta\sigma}\langle 0 |
u_\eta^a(0) s_\theta^b(x)
d_\phi^c(0) | \Sigma (p)\rangle~ ,\nonumber\\
\end{eqnarray}
\begin{eqnarray}\label{mut.mm}
\Pi^{II}_{\mu} &=& \frac{-i}{\sqrt{2}} \epsilon^{abc}\int d^4x
e^{-iqx} \Bigg\{\Bigg(\Big[( C )_{\eta\beta} (\gamma_5)_{\rho\phi}-(
C )_{\beta\phi} (\gamma_5)_{\rho\eta}\Big]  +\beta\Bigg[(C \gamma_5
)_{\eta\beta}(I)_{\rho\phi}
 \nonumber \\
&-& (C \gamma_5 )_{\beta\phi}(I)_{\rho\eta} \Bigg]\Bigg) \Big[
i\sigma_{\mu\nu}q^\nu(1+\gamma_5)
\Big]_{\sigma\theta}\Bigg\}S_Q(-x)_{\beta\sigma}  \langle 0 |
u_\eta^a(0) s_\theta^b(x)
d_\phi^c(0) | \Sigma (p)\rangle~ ,\nonumber\\
\end{eqnarray}

The heavy quark propagator, $ S_Q(x)$
is calculated in \cite{22Balitsky}:

\begin{eqnarray}\label{heavylightguy}
 S_Q (x)& =&  S_Q^{free} (x) - i g_s \int \frac{d^4 k}{(2\pi)^4}
e^{-ikx} \int_0^1 dv \Bigg[\frac{\not\!k + m_Q}{( m_Q^2-k^2)^2}
G^{\mu\nu}(vx) \sigma_{\mu\nu} \nnb \\
\ar \frac{1}{m_Q^2-k^2} v x_\mu G^{\mu\nu} \gamma_\nu \Bigg]~.
\end{eqnarray}
where,
\begin{eqnarray}\label{freeprop} S^{free}_{Q}
\es\frac{m_{Q}^{2}}{4\pi^{2}}\frac{K_{1}(m_{b}\sqrt{-x^2})}{\sqrt{-x^2}}-i
\frac{m_{Q}^{2}\not\!x}{4\pi^{2}x^2}K_{2}(m_{b}\sqrt{-x^2})~,
\end{eqnarray}
 and
$K_i$ are the Bessel functions. Note that $S_Q^{free}$ represents the free propagation of the heavy quark, and the remaining terms represent the interaction of the heavy
quark with the external gluon field. The calculation of the contributions of the latter effects require the four-  and five-particle baryons DA's which are currently unknown. But since
they are higher order contributions, they are expected to be small \cite{ek1,ek2,ek3} and we ignore them in the  present work. In \cite{ek4}, it is also found that the  form factors entering the semileptonic
decays of the heavy $\Lambda$ baryons turn
out to receive only a very small contribution from the gluon condensate.

The matrix element $\epsilon^{abc} \langle 0 |
u_\eta^a(0) s_\theta^b(x) d_\phi^c(0) | \Sigma (p)\rangle$ can be expressed in terms of $\Sigma$ baryon's wave functions and  are
given in  \cite{Liu}, and for completeness explicit form of them are
presented in the Appendix. After evaluating the Fourier transform, the correlation function is expressed in terms of the QCD parameters and the DA's of the $\Sigma$.

The sum rules are obtained by first Borel transforming both expression of the correlation functions and then equating the coefficient of various structures. Finally, the contributions of the higher states and the continuum are subtracted using quark hadron duality.
To extract the numerical value  of the form factors, value of the residue is also required. The residue of the $\Sigma_b$ baryon is calculated in \cite{Azizi-Bayar}.

\section{Numerical Analysis}
In this section, we perform  numerical analysis of the form factors
and use them to predict the decay rate and the branching ratio. The
masses of the $\Sigma_b$, $\Sigma$ baryons and the $b$ quark are
taken as $m_{\Sigma_{b}} = (5807.8 \pm2.7)~ MeV$ \cite{R311},
$m_{\Sigma} = (1192.642 \pm0.024)~ MeV$, and $m_b = (4.7\pm
0.1)~GeV$, respectively. For CKM matrix element entering into the
transition amplitude, $|V_{tb}V_{ts}^* |=0.041$ is used. The main
input parameters of QCD sum rules for the form factors are   DA's of
the $\Sigma$ baryon, whose explicit expression are presented in the
Appendix. Here, we should make the following remark. In the matrix
element, $\epsilon^{abc} \langle 0 | u_\eta^a(0) s_\theta^b(x)
d_\phi^c(0) | \Sigma (p)\rangle$, besides the functions presented in
the Appendix, there appear also the functions $\mathcal{A}_1^M$,
$\mathcal{V}_1^M$ and $\mathcal{T}_1^M$ whose explicit forms are
unknown for the $\Sigma$ baryon. Considering the $SU(3)_f$ symmetry,
we get them from the nucleon DA's. Our calculations show that their
contribution constitutes only few percent of the final results, so
we neglect their contribution in the present work.

%Four independent
%parameters contained in DA's are \cite{Liu}:
%\begin{eqnarray}
%f_{\Sigma}&=&(9.4\pm0.4)\times10^{-3}\; \mbox{GeV}^2,\hspace{2.5cm}\lambda_1=-%(2.5\pm0.1)\times10^{-2}\; \mbox{GeV}^2,\nonumber\\
%\lambda_2&=&(4.4\pm0.1)\times10^{-2}\;
%\mbox{GeV}^2,\hspace{2.5cm}\lambda_3=(2.0\pm0.1)\times10^{-2}\;
%\mbox{GeV}^2.\label{sigmapara}
%\end{eqnarray}

Besides these input parameters, there appear also three auxiliary
parameters in the sum rules, i.e. Borel mass parameter
$M^2$, continuum threshold $s_0$, and the general parameter
$\beta$ arising in the interpolating current of the  $\Sigma_b$ baryon. These parameters should not effect the values of the
form factors, so one should
obtain working regions of them for which the form factors show weak dependence on these parameters.

Lowering the value of the Borel mass increases the contribution of the higher twist DA's, hence requiring that the twist expansion converges leads to a lower limit on the
Borel mass; on the other hand, increasing the value of the Borel mass increases the contribution of the higher states and the continuum. Hence requiring that the contribution of the higher states and continuum to the correlation function is less than half the total contribution yields an upper bound on the Borel mass.
%\textit{The region for the Borel
%mass parameter is determined following that the higher states and
%continuum contributions makes up a small percentage of the total
%dispersion integral and the series on the light cone  in
%increasing twist should be convergent}. The former (latter) is used
%to determine the lower (upper) limit of Borel parameter and
Both of these conditions are met if the Borel mass is chosen in the interval
%the resultant interval is obtained as:
 $15~GeV^2\leq M_B^2\leq30~GeV^2$. The continuum threshold $s_0$ is not totally arbitrary but it is correlated to the energy
of the first excited state. Our analysis shows that in the region
$(m_{\Sigma_b}+0.3~GeV)^2\leq s_0\leq(m_{\Sigma_b}+0.7~GeV)^2$, the dependences of the form factors on this parameter are weak.
Finally, to find the working region of $\beta$, the dependence of the
form factors on $\cos \theta$ in the interval $-1\leq \cos \theta
\leq 1$, where $\tan \theta=\beta$ is considered. Our numerical calculations lead to the working region,
$-0.5\leq \cos\theta\leq0.7$.

\begin{figure}[htbp]
\subfigure[]{
\includegraphics[width=9cm]{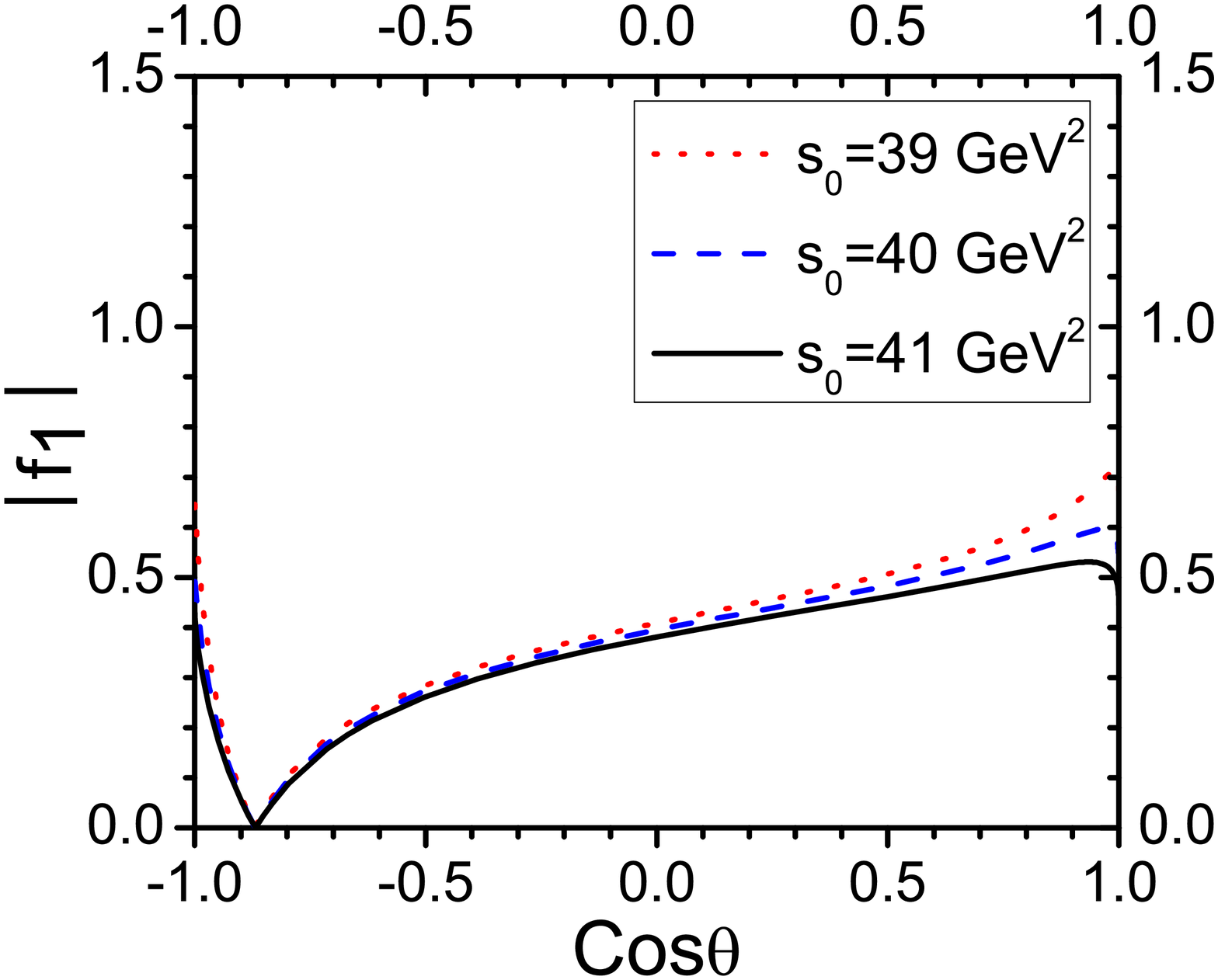}
}
\subfigure[]{
\includegraphics[width=9cm]{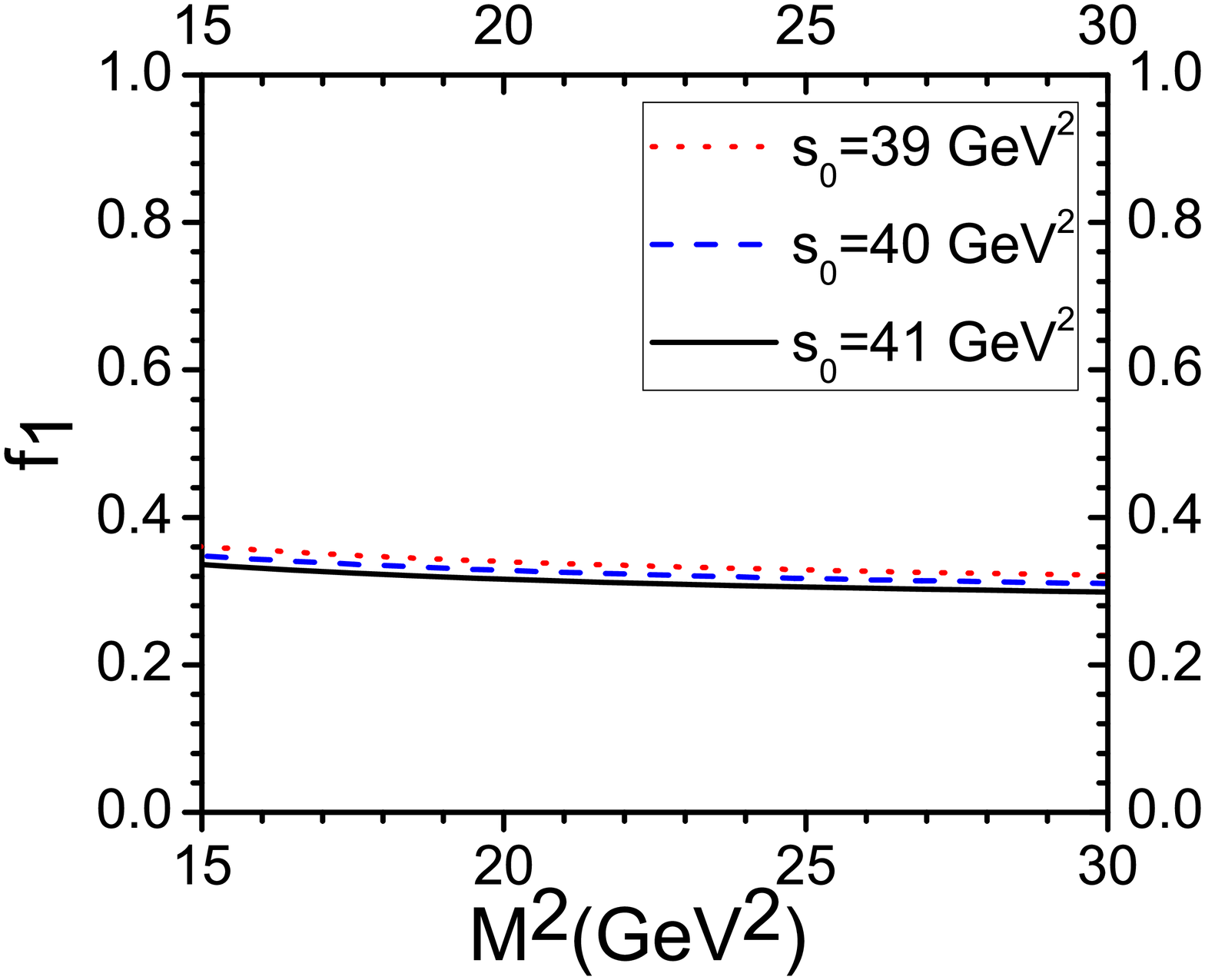}
}
\subfigure[]{
\includegraphics[width=9cm]{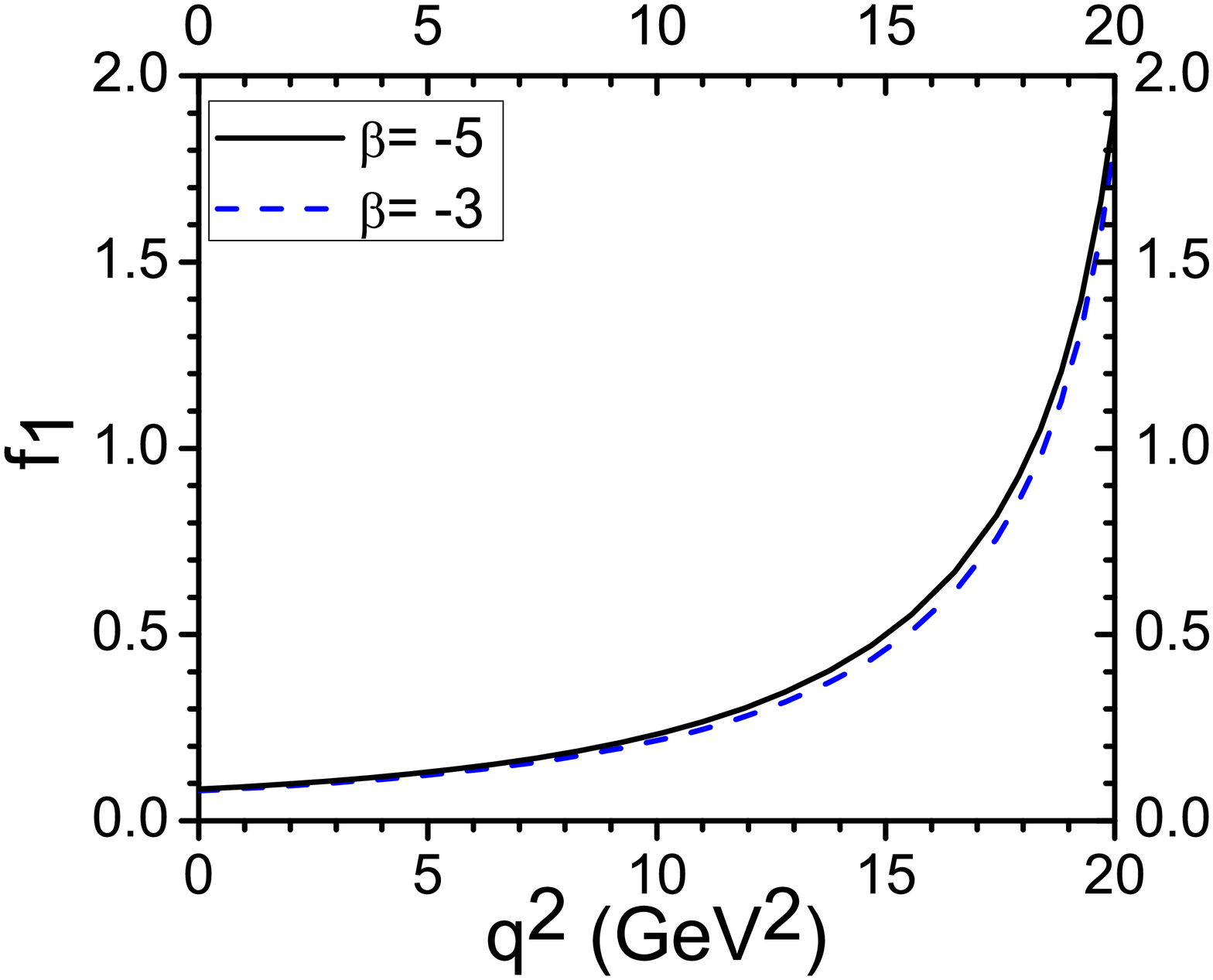}
}
\caption{Analysis of the sum rules for the form factor $f_1(q^2)$}
\label{fig1}
\end{figure}

As an example, in  Fig. 1,  we depict the dependences of  the form factor $f_1(q^2)$ on  auxiliary parameters as well as $q^2$.
Fig. (1a) shows the dependence of this form factor on $\cos \theta$ at the fixed values $q^2=13~GeV^2$, $M^2=20~GeV^2$ and $s_0=(40\pm1)~GeV^2$. As it is seen, there is a stable region in the interval
$-0.5\leq \cos\theta\leq0.7$. In Fig (1b), the Borel mass dependence of the same form factor is depicted at $q^2=13~GeV^2$ and the same value of the continuum threshold.
In the chosen working region of the  Borel mass, our predictions change by approximately $5\%$. From these two figures, it is also seen that our predictions are almost independent of the  continuum threshold.
 Finally, in Fig (1c), we show the dependence of the form factor $f_1$ on $q^2$ at two fixed values of $\beta$, and  $M^2=20~GeV^2$ and $s_0=40~GeV^2$.

The sum rules predictions are only reliable for the region $q^2 \ll m_b^2$, where in the decay of $\Sigma_b$, the allowed range of $q^2$ extends until $(m_{\Sigma_b} - m_\Sigma)^2$.
To extend the sum rules predictions to the whole physical region, the sum rules predictions are fitted to the following function:
\begin{eqnarray}
f_i(q^2)[g_i(q^2)]=\frac{a}{(1-\frac{q^2}{m_{fit}^2})}+\frac{b}{(1-\frac{q^2}{m_{fit}^2})^2}.
\label{parametrization1}
\end{eqnarray}
The central values of the fit parameters $a$, $b$, and $m_{fit}$ are
presented in Table~\ref{tab:13}. This table also exhibits  values of
the  form factors at $q^2=0$. These errors presented in this table
are due to  the variation of the auxiliary parameters, $M^2$, $s_0$,
and $\beta$, as well as the errors in the input parameters. Note
that for all form factors, the fit mass is always in the range
$m_{fit} = (5.1-5.4)~GeV$. In a vector dominance model, although the
double pole structure would not be expected, these form factors
would have poles at the masses of the (axial)vector meson that
couples to the transition currents. The observed (axial)vector $B_s$
mesons have masses in the range $(5.4-5.8)~GeV$. Although the fit
mass tends to be slightly smaller than the observed masses,
considering the uncertainties inherent in the sum rules
calculations, the results are reasonable. To improve the results one should consider the $\alpha_s$ corrections to the distributions amplitudes and
 more accurately determine the DA's of $\Sigma$ baryon.

Finally we calculate the differential and total decay rate of the $\Sigma_{b}\rar
\Sigma \ell^+\ell^-$ transition. The general form of the differential rate for the rare baryonic weak decay is given by
\cite{Aliev-Azizi} :

\begin{eqnarray} \frac{d\Gamma}{ds} = \frac{G^2\alpha^2_{em} m_{\Lambda_b}}{8192
\pi^5}| V_{tb}V_{ts}^*|^2 v \sqrt{\lambda} \, \Bigg[ \Theta(s) +
\frac{1}{3} \Delta(s)\Bigg]~, \label{rate} \end{eqnarray} where
$s=q^2/m^2_{\Sigma_b}$, $G = 1.17 \times 10^{-5}$ GeV$^{-2}$ is
the Fermi coupling constant and
$\lambda=\lambda(1, r, s)$ with $\lambda(a,b,c)=a^2+b^2+c^2-2ab-2ac-2bc$ is the usual triangle
function. Here, $v=\sqrt{1-\frac{4 m_\ell^2}{q^2}}$ is the lepton
velocity.  The functions $\Theta(s)$ and $\Delta(s)$ are given as:

\bea \Theta(s) \es 32 m_\ell^2 m_{\Sigma_b}^4 s (1+r-s) \ga \vel D_3
\ver^2 +
\vel E_3 \ver^2 \dr \nnb \\
\ar 64 m_\ell^2 m_{\Sigma_b}^3 (1-r-s) \, \mbox{\rm Re} [D_1^\ast
E_3 + D_3
E_1^\ast] \nnb \\
\ar 64 m_{\Sigma_b}^2 \sqrt{r} (6 m_\ell^2 - m_{\Sigma_b}^2 s)
{\rm Re} [D_1^\ast E_1] \nnb \\
\ar 64 m_\ell^2 m_{\Sigma_b}^3 \sqrt{r} \Big( 2 m_{\Sigma_b} s {\rm
Re} [D_3^\ast E_3] + (1 - r + s)
{\rm Re} [D_1^\ast D_3 + E_1^\ast E_3]\Big) \nnb \\
\ar 32 m_{\Sigma_b}^2 (2 m_\ell^2 + m_{\Sigma_b}^2 s) \Big\{ (1 - r
+ s) m_{\Sigma_b} \sqrt{r} \,
\mbox{\rm Re} [A_1^\ast A_2 + B_1^\ast B_2] \nnb \\
\ek m_{\Sigma_b} (1 - r - s) \, \mbox{\rm Re} [A_1^\ast B_2 +
A_2^\ast B_1] - 2 \sqrt{r} \Big( \mbox{\rm Re} [A_1^\ast B_1] +
m_{\Sigma_b}^2 s \,
\mbox{\rm Re} [A_2^\ast B_2] \Big) \Big\} \nnb \\
\ar 8 m_{\Sigma_b}^2 \Big\{ 4 m_\ell^2 (1 + r - s) + m_{\Sigma_b}^2
\Big[(1-r)^2 - s^2 \Big]
\Big\} \ga \vel A_1 \ver^2 +  \vel B_1 \ver^2 \dr \nnb \\
\ar 8 m_{\Sigma_b}^4 \Big\{ 4 m_\ell^2 \Big[ \lambda + (1 + r - s) s
\Big] + m_{\Sigma_b}^2 s \Big[(1-r)^2 - s^2 \Big]
\Big\} \ga \vel A_2 \ver^2 +  \vel B_2 \ver^2 \dr \nnb \\
\ek 8 m_{\Sigma_b}^2 \Big\{ 4 m_\ell^2 (1 + r - s) - m_{\Sigma_b}^2
\Big[(1-r)^2 - s^2 \Big]
\Big\} \ga \vel D_1 \ver^2 +  \vel E_1 \ver^2 \dr \nnb \\
\ar 8 m_{\Sigma_b}^5 s v^2 \Big\{ - 8 m_{\Sigma_b} s \sqrt{r}\,
\mbox{\rm Re} [D_2^\ast E_2] +
4 (1 - r + s) \sqrt{r} \, \mbox{\rm Re}[D_1^\ast D_2+E_1^\ast E_2]\nnb \\
\ek 4 (1 - r - s) \, \mbox{\rm Re}[D_1^\ast E_2+D_2^\ast E_1] +
m_{\Sigma_b} \Big[(1-r)^2 -s^2\Big] \ga \vel D_2 \ver^2 + \vel E_2
\ver^2\dr \Big\}~,
\eea \bea \Delta \left( s\right) \es - 8 m_{\Sigma_b}^4 v^2 \lambda
\ga \vel A_1 \ver^2 + \vel B_1 \ver^2 + \vel D_1 \ver^2
+ \vel E_1 \ver^2 \dr \nnb \\
\ar 8 m_{\Sigma_b}^6 s v^2 \lambda \Big( \vel A_2 \ver^2 + \vel B_2
\ver^2 + \vel D_2 \ver^2 + \vel E_2 \ver^2  \Big)~, \eea where  $r=
m^2_{\Sigma}/m^2_{\Sigma_b}$  and
 \bea \label{a9} A_1 \es
\frac{1}{q^2}\ga
f_1^T+g_1^T \dr \ga -2 m_b C_7\dr + \ga f_1-g_1 \dr C_9^{eff} \nnb \\
A_2 \es A_1 \ga 1 \rar 2 \dr ~,\nnb \\
A_3 \es A_1 \ga 1 \rar 3 \dr ~,\nnb \\
B_1 \es A_1 \ga g_1 \rar - g_1;~g_1^T \rar - g_1^T \dr ~,\nnb \\
B_2 \es B_1 \ga 1 \rar 2 \dr ~,\nnb \\
B_3 \es B_1 \ga 1 \rar 3 \dr ~,\nnb \\
D_1 \es \ga f_1-g_1 \dr C_{10} ~,\nnb \\
D_2 \es D_1 \ga 1 \rar 2 \dr ~, \\
D_3 \es D_1 \ga 1 \rar 3 \dr ~,\nnb \\
E_1 \es D_1 \ga g_1 \rar - g_1 \dr ~,\nnb \\
E_2 \es E_1 \ga 1 \rar 2 \dr ~,\nnb \\
E_3 \es E_1 \ga 1 \rar 3 \dr ~. \eea

Integrating the differential decay rate over   $s$ in the interval, $ 4m_\ell^2/m^2_{\Sigma_b} \leq s\leq (1- \sqrt{r})^2$,
we get the total  decay rates presented in the Table~\ref{tab:27}.
Finally, to obtain the branching ratios,
 one needs the lifetime of
the $\Sigma_b$ baryon. Although there is no exact information about the
lifetime of this baryon, it may be informative to take this lifetime
approximately at the same order of the $b$-baryon admixture,
($\Lambda_b, \Xi_b, \Sigma_b, \Omega_b$) which is
$\tau=1.391^{+0.039}_{-0.038}\times10^{-12}$~s \cite{R311}. The
results of the branching ratios for different leptons  are also presented in Table~\ref{tab:27}.
It is seen that the branching ratio for decays into the electrons or muons are more or less the same, while the branching ratio for
decay into final states containing $\tau$ lepton is reduced by approximately a factor of four. The order of branching fractions show that these channels can be detected at LHC.
%With the increment in the lepton masses, the results of decay rates
%decrease, and considering the corresponding phase space for each
%transition that is a reasonable consequence.
Comparing the presented results  in
this work with  the results of any
measurements, one can obtain useful information about the nature of
$\Sigma_b$ baryon as well as new physics effects beyond the SM.

\begin{table}[h]
\renewcommand{\arraystretch}{1.5}
\addtolength{\arraycolsep}{3pt}
$$
\begin{array}{|c|c|c|c|c|c|}

\hline \hline
                & \mbox{a} & \mbox{b}  & m_{fit}& q^2=0
                \\
\hline
 f_1            &  -0.035 &   0.13   &  5.1   &  0.095 \pm 0.017  \\
 f_2            &  0.026  &  -0.081  &  5.2   & -0.055 \pm 0.012  \\
 f_3            &   0.013 &  -0.065  &  5.3   & -0.052 \pm 0.016  \\
 g_1            &  -0.031 &   0.15   &  5.3   &  0.12 \pm 0.03    \\
 g_2            &  0.015  &  -0.040  &  5.3   & -0.025 \pm 0.008  \\
 g_3            &  0.012  &  -0.047  &  5.4   & -0.035 \pm 0.009  \\
 f_1^{T}        &  1.0    &   -1.0   &  5.4   &  0.0\pm0.0        \\
 f_2^{T}        &  -0.29  &   0.42   &  5.4   &  0.13 \pm 0.04    \\
 f_3^{T}        &  -0.24  &   0.41   &  5.4   &  0.17 \pm 0.05    \\
 g_1^{T}        &  0.45   &   -0.46  &  5.4   & -0.010\pm 0.003   \\
 g_2^{T}        &  0.031  &   0.055  &  5.4   &  0.086 \pm 0.024  \\
 g_3^{T}        &  -0.011 &  -0.18   &  5.4   & -0.19 \pm 0.06    \\
\hline \hline
\end{array}
$$
\caption{Parameters appearing in  the fit function of the  form
factors, $f_{1}$, $f_{2}$, $f_{3}$, $g_{1}$, $g_{2}$, $g_{3}$,
$f^T_{1}$, $f^T_{2}$, $f^T_{3}$, $g^T_{1}$, $g^T_{2}$ and $g^T_{3}$
in full theory for $\Sigma_{b}\rightarrow \Sigma\ell^{+}\ell^{-}$
and the values of the form factors at $q^2=0$. In this Table only
central values of the parameters are presented.} \label{tab:13}
\renewcommand{\arraystretch}{1}
\addtolength{\arraycolsep}{-1.0pt}
\end{table}

\begin{table}[h] \centering
$$
\begin{array}{|c|c|c|c|c|}

\hline \hline
    & \Gamma(GeV)  & BR \\

\hline \Sigma_{b} \rar \Sigma e^+e^-& (4.30 \pm 0.82) \times
10^{-18} &(9.09\pm1.73)\times 10^{-6}\\ \hline \Sigma_{b} \rar
\Sigma \mu^+\mu^- & (4.29 \pm 0.82) \times
10^{-18}&(9.06\pm1.72)\times 10^{-6}
\\ \hline
\Sigma_{b}\rar \Sigma \tau^+\tau^- & (1.30 \pm 0.42) \times
10^{-18}&(2.75\pm0.88)\times 10^{-6}
\\ \hline

\end{array}
$$
\vspace{0.8cm} \caption{The values of the decay rate and branching
ratios for $\Sigma_{b}\rightarrow \Sigma\ell^{+}\ell^{-}$   for different leptons. In the case of decays into and electron positron pair, a lower cut-off of $q^2 \ge 0.04 ~GeV^2$ is imposed to avoid the resonance due to a real photon creating the electron-positron pair.} \label{tab:27}
\end{table}

\section{Acknowledgment}
This work is suppoerted by TUBITAK under the project No. 110T284.
M. Bayar also acknowledges support through TUBITAK BIDEP-2219 grant.

\newpage

\section*{Appendix A}

In this Appendix,  the general decomposition of the matrix element, $
\epsilon^{abc}\langle 0 |  u_\eta^a(0) d_\theta^b(x) s_\phi^c(0) |
\Sigma (p)\rangle$
 as well as the DA's of $\Sigma$ \cite{Liu} are presented.
Considering Lorentz and parity invariances, the matrix element can be decomposed into various Lorentz structures as:
\begin{eqnarray}\label{wave func}
&&4\langle0|\epsilon^{abc}u_\alpha^a(a_1 x)s_\beta^b(a_2
x)d_\gamma^c(a_3 x)|\Sigma(p)\rangle\nnb\\
\es\mathcal{S}_1m_{\Sigma}C_{\alpha\beta}(\gamma_5\Sigma)_{\gamma}+
\mathcal{S}_2m_{\Sigma}^2C_{\alpha\beta}(\rlap/x\gamma_5\Sigma)_{\gamma}\nnb\\
\ar \mathcal{P}_1m_{\Sigma}(\gamma_5C)_{\alpha\beta}\Sigma_{\gamma}+
\mathcal{P}_2m_{\Sigma}^2(\gamma_5C)_{\alpha\beta}(\rlap/x\Sigma)_{\gamma}+
(\mathcal{V}_1+\frac{x^2m_{\Sigma}^2}{4}\mathcal{V}_1^M)(\rlap/pC)_{\alpha\beta}(\gamma_5\Sigma)_{\gamma}
\nnb\\\ar
\mathcal{V}_2m_{\Sigma}(\rlap/pC)_{\alpha\beta}(\rlap/x\gamma_5\Sigma)_{\gamma}+
\mathcal{V}_3m_{\Sigma}(\gamma_\mu
C)_{\alpha\beta}(\gamma^\mu\gamma_5\Sigma)_{\gamma}+
\mathcal{V}_4m_{\Sigma}^2(\rlap/xC)_{\alpha\beta}(\gamma_5\Sigma)_{\gamma}\nnb\\\ar
\mathcal{V}_5m_{\Sigma}^2(\gamma_\mu
C)_{\alpha\beta}(i\sigma^{\mu\nu}x_\nu\gamma_5\Sigma)_{\gamma} +
\mathcal{V}_6m_{\Sigma}^3(\rlap/xC)_{\alpha\beta}(\rlap/x\gamma_5\Sigma)_{\gamma}
+(\mathcal{A}_1\nnb\\
\ar\frac{x^2m_{\Sigma}^2}{4}\mathcal{A}_1^M)(\rlap/p\gamma_5
C)_{\alpha\beta}\Sigma_{\gamma}+
\mathcal{A}_2m_{\Sigma}(\rlap/p\gamma_5C)_{\alpha\beta}(\rlap/x\Sigma)_{\gamma}+
\mathcal{A}_3m_{\Sigma}(\gamma_\mu\gamma_5
C)_{\alpha\beta}(\gamma^\mu \Sigma)_{\gamma}\nnb\\\ar
\mathcal{A}_4m_{\Sigma}^2(\rlap/x\gamma_5C)_{\alpha\beta}\Sigma_{\gamma}+
\mathcal{A}_5m_{\Sigma}^2(\gamma_\mu\gamma_5
C)_{\alpha\beta}(i\sigma^{\mu\nu}x_\nu \Sigma)_{\gamma}+
\mathcal{A}_6m_{\Sigma}^3(\rlap/x\gamma_5C)_{\alpha\beta}(\rlap/x
\Sigma)_{\gamma}\nnb\\\ar(\mathcal{T}_1+\frac{x^2m_{\Sigma}^2}{4}\mathcal{T}_1^M)(p^\nu
i\sigma_{\mu\nu}C)_{\alpha\beta}(\gamma^\mu\gamma_5
\Sigma)_{\gamma}+\mathcal{T}_2m_{\Sigma}(x^\mu p^\nu
i\sigma_{\mu\nu}C)_{\alpha\beta}(\gamma_5 \Sigma)_{\gamma}\nnb\\\ar
\mathcal{T}_3m_{\Sigma}(\sigma_{\mu\nu}C)_{\alpha\beta}(\sigma^{\mu\nu}\gamma_5
\Sigma)_{\gamma}+
\mathcal{T}_4m_{\Sigma}(p^\nu\sigma_{\mu\nu}C)_{\alpha\beta}(\sigma^{\mu\rho}x_\rho\gamma_5
\Sigma)_{\gamma}\nnb\\\ar \mathcal{T}_5m_{\Sigma}^2(x^\nu
i\sigma_{\mu\nu}C)_{\alpha\beta}(\gamma^\mu\gamma_5
\Sigma)_{\gamma}+ \mathcal{T}_6m_{\Sigma}^2(x^\mu p^\nu
i\sigma_{\mu\nu}C)_{\alpha\beta}(\rlap/x\gamma_5
\Sigma)_{\gamma}\nnb\\
\ar
\mathcal{T}_7m_{\Sigma}^2(\sigma_{\mu\nu}C)_{\alpha\beta}(\sigma^{\mu\nu}\rlap/x\gamma_5
\Sigma)_{\gamma}+
\mathcal{T}_8m_{\Sigma}^3(x^\nu\sigma_{\mu\nu}C)_{\alpha\beta}(\sigma^{\mu\rho}x_\rho\gamma_5
\Sigma)_{\gamma}~.\nnb~~~~~~~~~~~~~~~~~~~~~~~(A.1) \end{eqnarray}

The calligraphic functions in the above expression  do not have
definite twists but they can be written in terms of the $\Sigma$
distribution amplitudes (DA's) with definite and  increasing twists
via   the scalar product $px$ and the parameters $a_i$, $i=1,2,3$.
The relationship between the calligraphic functions appearing in the
above equation and scalar, pseudo-scalar, vector, axial vector and
tensor DA's for $\Sigma$ are given in Tables \ref{tab:1}, \ref{tab:2},
\ref{tab:3}, \ref{tab:4} and \ref{tab:5}, respectively.
\begin{table}[h]
\centering
\begin{tabular}{|c|} \hline
$\mathcal{S}_1 = S_1$\\ \hline\hline
 $2px\mathcal{S}_2=S_1-S_2$ \\ \hline
   \end{tabular}
\vspace{0.3cm} \caption{Relations between the calligraphic functions
and $\Sigma$ scalar DA's.}\label{tab:1}
\end{table}
\begin{table}[h]
\centering
\begin{tabular}{|c|} \hline
  $\mathcal{P}_1=P_1$\\ \hline
  $2px\mathcal{P}_2=P_1-P_2$ \\ \hline
   \end{tabular}
\vspace{0.3cm} \caption{Relations between the calligraphic functions
and $\Sigma$ pseudo-scalar DA's.}\label{tab:2}
\end{table}
\begin{table}[h]
\centering
\begin{tabular}{|c|} \hline
  $\mathcal{V}_1=V_1$ \\ \hline
  $2px\mathcal{V}_2=V_1-V_2-V_3$ \\ \hline
  $2\mathcal{V}_3=V_3$ \\ \hline
  $4px\mathcal{V}_4=-2V_1+V_3+V_4+2V_5$ \\ \hline
  $4px\mathcal{V}_5=V_4-V_3$ \\ \hline
  $4(px)^2\mathcal{V}_6=-V_1+V_2+V_3+V_4
 + V_5-V_6$ \\ \hline
 \end{tabular}
\vspace{0.3cm} \caption{Relations between the calligraphic functions
and $\Sigma$ vector DA's.}\label{tab:3}
\end{table}
\begin{table}[h]
\centering
\begin{tabular}{|c|} \hline
  $\mathcal{A}_1=A_1$ \\ \hline
  $2px\mathcal{A}_2=-A_1+A_2-A_3$ \\ \hline
   $2\mathcal{A}_3=A_3$ \\ \hline
  $4px\mathcal{A}_4=-2A_1-A_3-A_4+2A_5$ \\ \hline
  $4px\mathcal{A}_5=A_3-A_4$ \\ \hline
  $4(px)^2\mathcal{A}_6=A_1-A_2+A_3+A_4-A_5+A_6$ \\ \hline
 \end{tabular}
\vspace{0.3cm} \caption{Relations between the calligraphic functions
and $\Sigma$ axial vector DA's.}\label{tab:4}
\end{table}
\begin{table}[h]
\centering
\begin{tabular}{|c|} \hline
  $\mathcal{T}_1=T_1$ \\ \hline
  $2px\mathcal{T}_2=T_1+T_2-2T_3$ \\ \hline
  $2\mathcal{T}_3=T_7$ \\ \hline
  $2px\mathcal{T}_4=T_1-T_2-2T_7$ \\ \hline
  $2px\mathcal{T}_5=-T_1+T_5+2T_8$ \\ \hline
  $4(px)^2\mathcal{T}_6=2T_2-2T_3-2T_4+2T_5+2T_7+2T_8$ \\ \hline
  $4px \mathcal{T}_7=T_7-T_8$\\ \hline
  $4(px)^2\mathcal{T}_8=-T_1+T_2 +T_5-T_6+2T_7+2T_8$\\ \hline
 \end{tabular}
\vspace{0.3cm} \caption{Relations between the calligraphic functions
and $\Sigma$ tensor DA's.}\label{tab:5}
\end{table}

Every distribution amplitude $F(a_ipx)$=  $S_i$, $P_i$, $V_i$,
$A_i$, $T_i$ can be represented as:
\begin{eqnarray}\label{dependent1} F(a_ipx)=\int
dx_1dx_2dx_3\delta(x_1+x_2+x_3-1) e^{-ip
x\sum_ix_ia_i}F(x_i)~.\nnb~~~~~~~~~~~~~~~~~~~~~~~~~~~~~~(A.2)
\end{eqnarray}
where, $x_{i}$ with $i=1,~2$ and $3$ are longitudinal momentum
fractions carried by the participating quarks.

The explicit expressions for the  $\Sigma$ DA's up to twists 6 are
given as follows \cite{Liu}:

Twist-$3$ distribution amplitudes:
\begin{eqnarray}
V_1(x_i)&=&120x_1x_2x_3\phi_3^0\,,\hspace{2.5cm}A_1(x_i)=0\,,\nonumber\\
T_1(x_i)&=&120x_1x_2x_3\phi_3^{'0}\,.
\end{eqnarray}
Twist-$4$ distribution amplitudes:
\begin{eqnarray}
S_1(x_i)&=&6(x_2-x_1)x_3(\xi_4^0+\xi_4^{'0})\,,\hspace{1.6cm}P_1(x_i)=6(x_2-x_1)x_3(\xi_4^0-\xi_4^{'0})\,,\nonumber\\
V_2(x_i)&=&24x_1x_2\phi_4^0\,,\hspace{3.9cm}A_2(x_i)=0\,,\nonumber\\
V_3(x_i)&=&12x_3(1-x_3)\psi_4^0\,,\hspace{2.8cm}A_3(x_i)=-12x_3(x_1-x_2)\psi_4^0\,,\nonumber\\
T_2(x_i)&=&24x_1x_2\phi_4^{'0}\,,\hspace{3.9cm}T_3(x_i)=6x_3(1-x_3)(\xi_4^0+\xi_4^{'0})\,,\nonumber\\
T_7(x_i)&=&6x_3(1-x_3)(\xi_4^{'0}-\xi_4^0)\,.
\end{eqnarray}
Twist-$5$ distribution amplitudes:
\begin{eqnarray}
S_2(x_i)&=&\frac32(x_1-x_2)(\xi_5^0+\xi_5^{'0})\,,\hspace{1.5cm}P_2(x_i)=\frac32(x_1-x_2)(\xi_5^0-\xi_5^{'0})\,,\nonumber\\
V_4(x_i)&=&3(1-x_3)\psi_5^0\,,\hspace{2.95cm}A_4(x_i)=3(x_1-x_2)\psi_5^0\,,\nonumber\\
V_5(x_i)&=&6x_3\phi_5^0\,,\hspace{4.05cm}A_5(x_i)=0\,,\nonumber\\
T_4(x_i)&=&-\frac32(x_1+x_2)(\xi_5^{'0}+\xi_5^0)\,,\hspace{1.25cm}T_5(x_i)=6x_3\phi_5^{'0}\,,\nonumber\\
T_8(x_i)&=&\frac32(x_1+x_2)(\xi_5^{'0}-\xi_5^0)\,.
\end{eqnarray}
Finally, twist-$6$ distribution amplitudes:
\begin{eqnarray}
V_6(x_i)&=&2\phi_6^0\,,\hspace{2.5cm}A_6(x_i)=0\,,\nonumber\\
T_6(x_i)&=&2\phi_6^{'0}\,.
\end{eqnarray}

\begin{eqnarray}
\phi_3^0&=&\phi_6^0=f_{\Sigma^+},\hspace{2.8cm}\psi_4^0=\psi_5^0=\frac12(f_{\Sigma^+}-\lambda_1)\,,\nonumber\\
\phi_4^0&=&\phi_5^0=\frac12(f_{\Sigma^+}+\lambda_1),\hspace{1.3cm}\phi_3'^0=\phi_6'^0=-\xi_5^0=\frac16(4\lambda_3-\lambda_2)\,,\nonumber\\
\phi_4'^0&=&\xi_4^0=\frac16(8\lambda_3-3\lambda_2),\hspace{1.2cm}\phi_5'^0=-\xi_5'^0=\frac16\lambda_2\,,\nonumber\\
\xi_4'^0&=&\frac16(12\lambda_3-5\lambda_2)\,,
\end{eqnarray}
where,
\begin{eqnarray}
f_{\Sigma}&=&(9.4\pm0.4)\times10^{-3}\; \mbox{GeV}^2,\hspace{2.5cm}
\lambda_1=-(2.5\pm0.1)\times10^{-2}\; \mbox{GeV}^2,\nonumber\\
\lambda_2&=&(4.4\pm0.1)\times10^{-2}\;
\mbox{GeV}^2,\hspace{2.5cm}\lambda_3=(2.0\pm0.1)\times10^{-2}\;
\mbox{GeV}^2.\label{sigmapara}
\end{eqnarray}
\end{document}